\documentclass[pre,twocolumn,aps]{revtex4}
\newcommand{\br}{{\bf r}}

\newcommand{\bR}{{\bf R}}
\newcommand{\bA}{{\bf A}}
\newcommand{\bF}{{\bf F}}

\newcommand{\bt}{{\bf t}}

\newcommand{\dt}{\Delta t}

\newcommand{\bD}{{\bm\Delta}}
\newcommand{\am}{\langle m\rangle}

\usepackage{graphicx}
\usepackage{comment}
\usepackage{bm}
\usepackage{amsmath,amssymb} 
\usepackage{epstopdf}
\usepackage{verbatim}
\usepackage{float}
\usepackage[font=small,format=plain,labelfont=bf,textfont=sl]{caption}

\begin{document}

\title{$NVU$ dynamics. II. Comparing to four other dynamics}
\author{Trond S. Ingebrigtsen, S{\o}ren Toxvaerd, Thomas B. Schr{\o}der, and Jeppe C. Dyre}
\email{dyre@ruc.dk}
\affiliation{DNRF Centre ``Glass and Time'', IMFUFA, Department of Sciences, Roskilde University, Postbox 260, DK-4000 Roskilde, Denmark}
\date{\today}

\begin{abstract}
In the companion paper [Ingebrigtsen \textit{et al.}] an algorithm was developed for tracing out a geodesic curve on the constant-potential-energy hypersurface. Here simulations of this $NVU$ dynamics are compared to results for four other dynamics, both deterministic and stochastic. First, $NVU$ dynamics is compared to the standard energy-conserving Newtonian $NVE$ dynamics by simulations of the Kob-Andersen binary Lennard-Jones liquid, its WCA version (i.e., with cut-off's at the pair potential minima), and the Gaussian Lennard-Jones liquid. We find identical results for all quantities probed: radial distribution functions, incoherent intermediate scattering functions, and mean-square displacement as function of time. Arguments are then presented for the equivalence of $NVU$ and $NVE$ dynamics in the thermodynamic limit; in particular to leading order in $1/N$ these two dynamics give identical time-autocorrelation functions. In the final section {\it NVU} dynamics is compared to Monte Carlo dynamics, to a diffusive dynamics of small-step random walks on the constant-potential-energy hypersurface, and to Nos$\acute{e}$-Hoover $NVT$ dynamics. If time is scaled for the two stochastic dynamics to make their single-particle diffusion constants identical to those of $NVE$ dynamics, the simulations show that all five dynamics are equivalent at low temperatures except at short times.
\end{abstract}

\pacs{64.70.Pf}

\maketitle

\section{Introduction}

In the companion paper (Paper I \cite{I}) we developed a stable numerical algorithm for tracing out a geodesic curve on the constant-potential-energy hypersurface $\Omega$ of a system of $N$ classical particles. If $U(\br_1,...,\br_N)$ is the potential energy as a function of the particle coordinates, for a given value $U_0$ of the potential energy $\Omega$ is the $3N-1$ dimensional Riemannian differentiable manifold defined by (where $\bR\equiv (\br_1,...,\br_N)$ is the position in the $3N$ dimensional configuration space)

\begin{equation}\label{omega_def}
  \Omega\,=\,\{\bR\in R^{3N} \,|\, U(\bR)\,=\,U_0\}\,.
\end{equation} 
Geodesic motion on $\Omega$ is termed $NVU$ dynamics in analogy with standard Newtonian $NVE$ dynamics, which conserves the total energy $E$. Motivations for studying $NVU$ dynamics were given in Paper I. The present paper compares $NVU$ dynamics to four other dynamics, two deterministic and two stochastic, concluding that $NVU$ dynamics is a fully valid molecular dynamics.

The path of shortest distance between two points on a Riemannian manifold is a so-called geodesic curve. By definition a geodesic is as a curve of stationary length, i.e., one for which small curve variations keeping the two end points $\bR_A$ and $\bR_B$ fixed to lowest order do not change the curve length, i.e.,

\begin{equation}\label{cond}
  \delta \int_{\bR_A}^{\bR_B} dl \,=\, 0\,.
\end{equation}
By discretizing this condition and carrying out the variation, keeping the potential energy fixed by introducing Lagrangian multipliers, a ``basic NVU algorithm'' was derived in Paper I ($\bF$ is the $3N$-dimensional force vector and $i$ is the time-step index):

\begin{equation}\label{numgeo}
  \bR_{i+1} \,=\, 2\bR_i - \bR_{i-1} -2\, \frac{\bF_{i} \cdot (\bR_{i} - \bR_{i-1})}{\bF_{i}^{2}} \bF_i\,.
\end{equation}
This algorithm works well, but for very long simulations numerical errors accumulate and $U$ drifts to higher values (``entropic drift'', see Paper I). This problem is also experienced for the total energy in $NVE$ algorithms \cite{sim}, and it is not more severe for $NVU$ than for $NVE$ dynamics. A fully stable $NVU$ algorithm was developed in Paper I, which may be summarized as follows. If one switches to the leap-frog representation and defines the position changes by $\bD_{i+1/2}=\bR_{i+1}-\bR_{i}$, the stable $NVU$ algorithm is: $ \bD_{i+1/2} =  l_0\,\bA_{i-1/2}/|\bA_{i-1/2}|$ where $l_0$ is the step length and $\bA_{i-1/2}=\bD_{i-1/2}+(-2\bF_i\cdot\bD_{i-1/2}+U_{i-1}-U_0){\bF_i}/{\bF_i^2}$. Just as for standard $NVE$ dynamics a final stabilization introduced is to adjust the position changes slightly, e.g., every 100th step, in order to eliminate numerical drift of the center of mass coordinate. In the simulations reported below we used the stabilized $NVU$ algorithm. However, since the stabilization is merely a technicality, the basic $NVU$ algorithm Eq. (\ref{numgeo}) is used for theoretical considerations. 

Constant-potential-energy algorithms were previously considered in papers dating back to 1986 by Cotterill and Madsen {\it et al.} \cite{cott} and in 2002 by Scala {\it et al.} \cite{scala}. In the same spirit, but a slightly different context, Stratt and coworkers in 2007 and 2010 considered geodesic motion in the space of points with potential energy lower than $U_0$ \cite{stratt}. In the thermodynamic limit these points are almost all of energy very close to $U_0$. We refer to Paper I for a more detailed discussion of how $NVU$ dynamics relates to these earlier works. 

$NVU$ dynamics invites to an alternative view of molecular motion. Instead of focusing on the standard potential-energy landscape in $(3N+1)$ dimensions \cite{landscape}, $NVU$ dynamics adopts the configuration-space microcanonical viewpoint and focuses on the $(3N-1)$-dimensional Riemannian hypersurface $\Omega$. The classical potential-energy landscape picture draws attention to the stationary points of the potential-energy function, in particular its minima, the so-called inherent states \cite{landscape}. In contrast, all points on $\Omega$ have the same probability in $NVU$ dynamics and there are no energy barriers -- all barriers are of entropic nature defining unlikely parts of $\Omega$ that must be passed  \cite{cott,scala,stratt}. Despite the absence of energy barriers in the ordinary sense of this term, $NVU$ dynamics is fully able to describe locally activated events (hopping processes between local potential-energy minima). The $NVU$ ``heat bath'' is provided by the multitude of configurational degrees of freedom \cite{cott,scala,stratt}.

The present paper compares $NVU$ dynamics to other molecular dynamics, including stochastic ones. We first compare to $NVE$ dynamics, which is also deterministic, and conclude that for large systems the two dynamics are equivalent. We proceed to compare to other kinds of dynamics, inspired by previous works: The first investigation providing long-time simulations that compared different dynamics (Newtonian versus Langevin) was presented by Gleim \textit{et al.} \cite{gleim}. They studied the Kob-Andersen binary Lennard-Jones (KABLJ) mixture \cite{ka} at different temperatures and found that below a certain temperature ($T < 0.8$), the temperature dependence of the diffusion constant and of the structural relaxation time was identical in the two dynamics. This type of investigation was extended by Szamel \textit{et al.} \cite{szamel} to Brownian dynamics, i.e., stochastic dynamics without the momentum degrees of freedom. They found power-law fitting exponents for the temperature dependence of the diffusion constant and relaxation time very close to those of $NVE$ dynamics. Subsequently Berthier \textit{et al.} \cite{berthierMC} investigated Monte Carlo dynamics for which agreement with Newtonian dynamics was also established, both for a strong and a fragile model glass former (an SiO$_{2}$ model and the KABLJ model). This, however, did not apply for higher-order time-correlation functions, a fact contributed to the presence of different conservation laws \cite{berthierMC}.  

We compare below $NVU$ dynamics to four other dynamics: Newtonian dynamics ($NVE$), Nos$\acute{e}$-Hoover \textit{NVT} dynamics \cite{nosehoovernvt}, Monte Carlo dynamics (\textit{MC}) \cite{metropolismc}, and a diffusive small-step random-walk dynamics on the potential energy hypersurface ($RW$). Section \ref{nvunve} compares $NVU$ dynamics with the ''true'' ($NVE$) time evolution defined by Newton's second law. This is done by simulations of the KABLJ liquid, as well as of the Weeks-Chandler-Andersen (WCA) approximation \cite{wca} to the KABLJ liquid (KABWCA). Section \ref{proof} gives some intuitive arguments for the equivalence of $NVU$ and $NVE$ dynamics in the thermodynamic limit.  Section \ref{nvuother} compares $NVU$ dynamics with \textit{NVT}, \textit{MC}, and $RW$ dynamics. Section \ref{sum} gives a brief summary and outlook.

\section{Simulations comparing $NVU$ dynamics to $NVE$ dynamics}\label{nvunve}

\begin{figure}[H]
  \includegraphics[width=70mm]{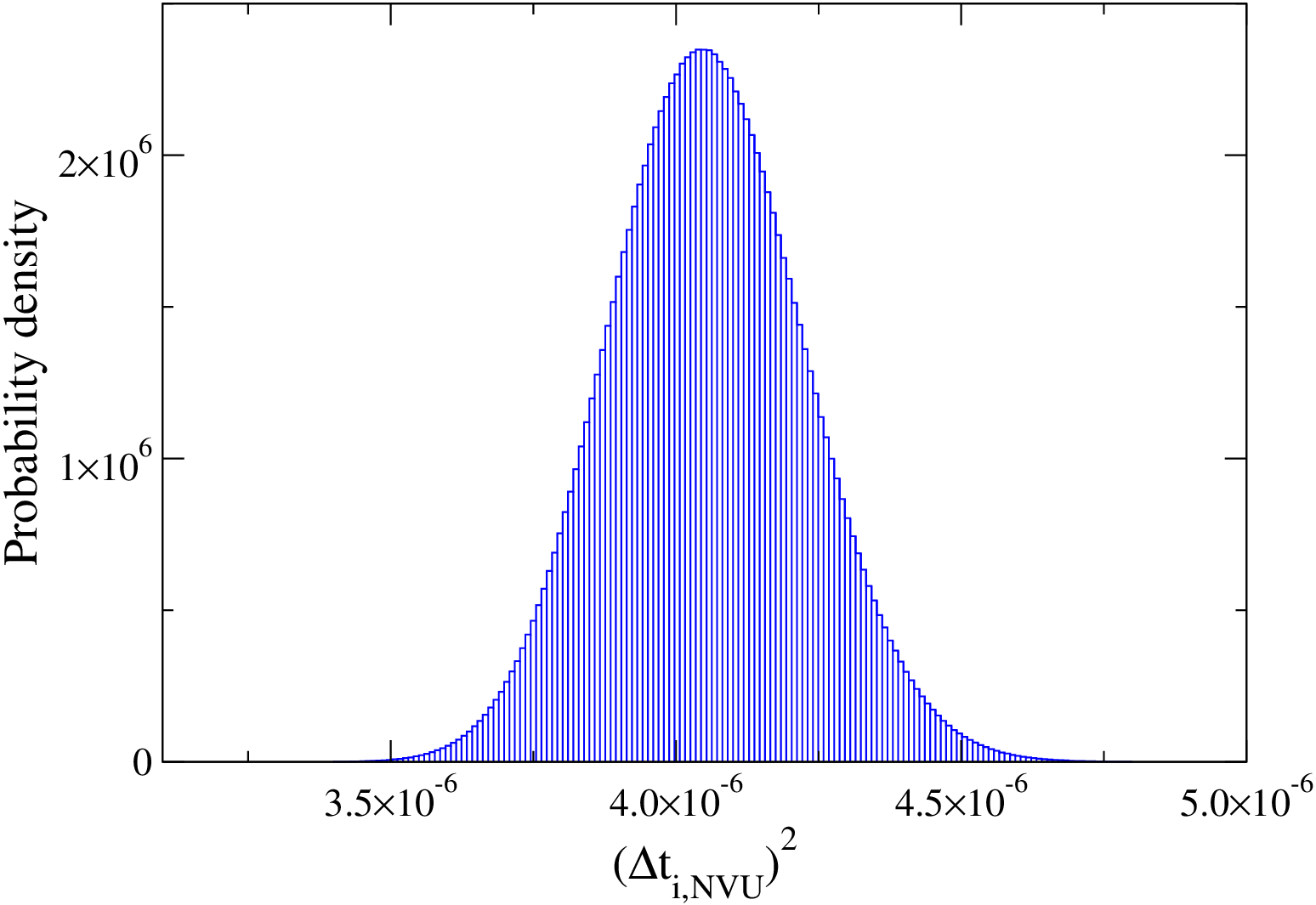}
  \includegraphics[width=70mm]{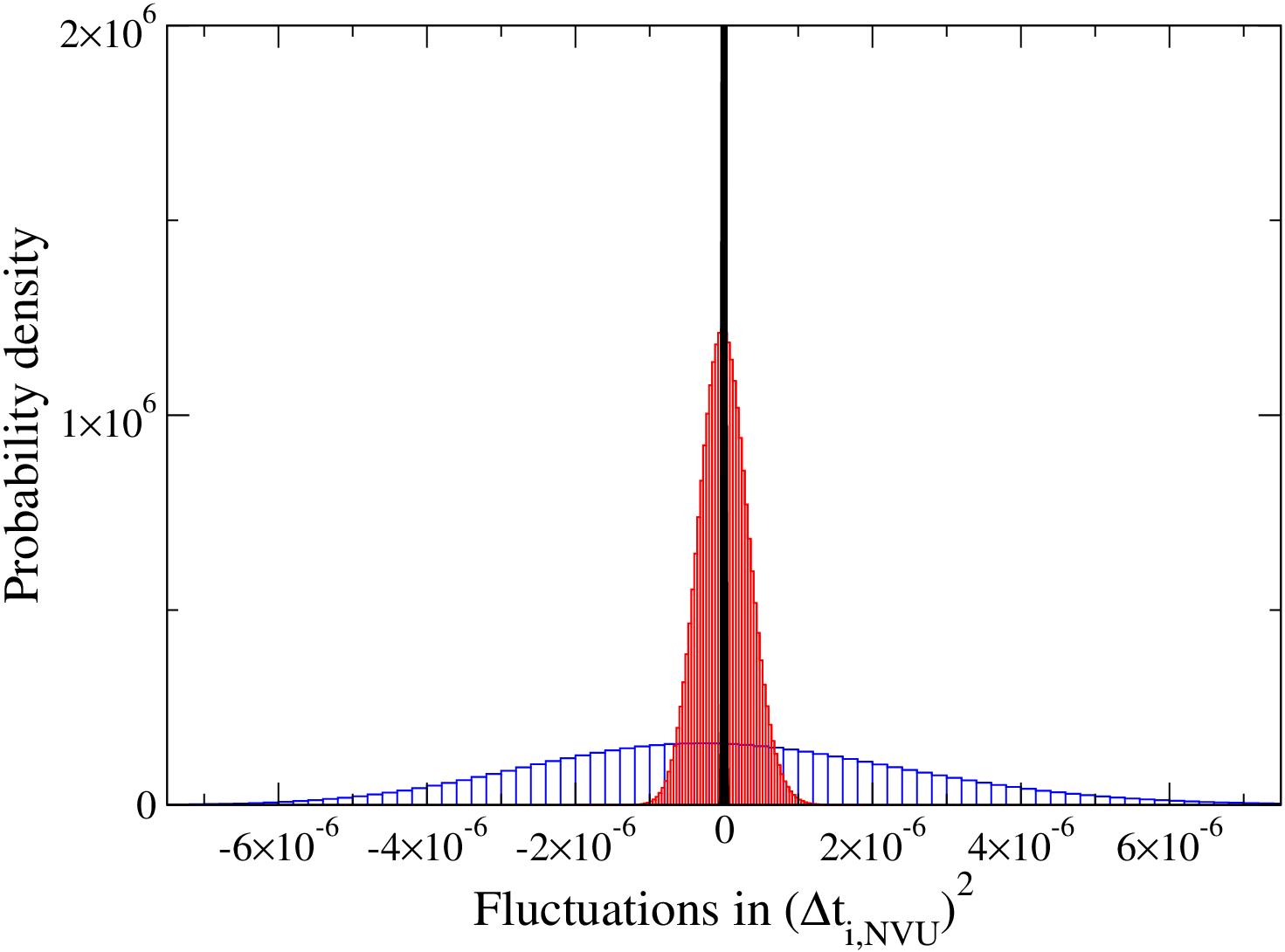}
  \caption{\label{ldist}
(a) Probability density of $(\dt_{i,NVU})^{2}$ given by Eq. (\ref{time_step}) for the Kob-Andersen binary Lennard-Jones (KABLJ) mixture at $\rho=1.2$ and $T=0.44$
(b) Probability density for  $(\dt_{i,NVU})^{2}-\langle(\dt_{i,NVU})^{2}\rangle$ for 256, 1024, and 8192 particles of the single-component LJ liquid ($T=0.70$, $\rho=0.85$), showing a narrowing as the particle number increases.
}
  \end{figure}

In $NVU$ dynamics a geodesic is traced in configuration space. Physically, this curve may be traversed with any velocity; comparing however to {\it NVE} dynamics suggests an obvious time measure for $NVU$ dynamics, as we shall see now. Limiting ourselves for simplicity to systems of particles with identical masses $m$, the Verlet algorithm for $NVE$ dynamics with time step $\dt_{NVE}$ is \cite{sim,verlet}

\begin{equation}\label{verlet}
\bR_{i+1} \,=
\, 2\,\bR_i - \bR_{i-1} +\frac{(\dt_{NVE})^2}{m} \bF_i\,.
\end{equation}
Comparing to Eq. (\ref{numgeo}) suggests the following identification of an $NVU$ time step $\dt_{i, NVU}$ 

\begin{equation}\label{time_step}
\frac{(\dt_{i,NVU})^2}{m}\,=\,  
-2\,\frac{\bF_{i} \cdot (\bR_{i} - \bR_{i-1})}{\bF_{i}^{2}}\,.
\end{equation}
This quantity is identical to $l_0\lambda$ of Paper I; our simulations show that the right-hand side is always positive for small $l_0$. We have no proof of this, but presumably it applies rigorously in the thermodynamic limit.

In the following data are given in terms of the natural units for the Lennard-Jones pair potential; for the KABLJ and KAWCA system length and energy are given in units of the large-particle parameters $\sigma_{AA}$ and $\epsilon_{AA}$, respectively. The simulation sizes are $N =1024, 1000$, and $1024$ for KABLJ, KAWCA, and Lennard-Jones Gaussian, respectively (see below). 

The probability distribution of $(\dt_{i,NVU})^{2}$ is given in Fig. \ref{ldist} for an $N=1024$ KABLJ liquid at $\rho=1.2$ and $T=0.44$ \cite{simulations}. The simulations behind this, as well as all below figures, were initiated by choosing the two initial configurations from a well-equilibrated $NVE$ simulation. The potential energy $U$ in the $NVU$ simulation was chosen as $U = \langle U\rangle_{NVE}$ at the relevant state points. The probability distribution of Fig. \ref{ldist} is a Gaussian, which is consistent with the fact that $(\dt_{i,NVU})^{2}$ is a sum of many terms that are uncorrelated for large spatial separations.

\begin{figure}[H]
    \includegraphics[width=70mm]{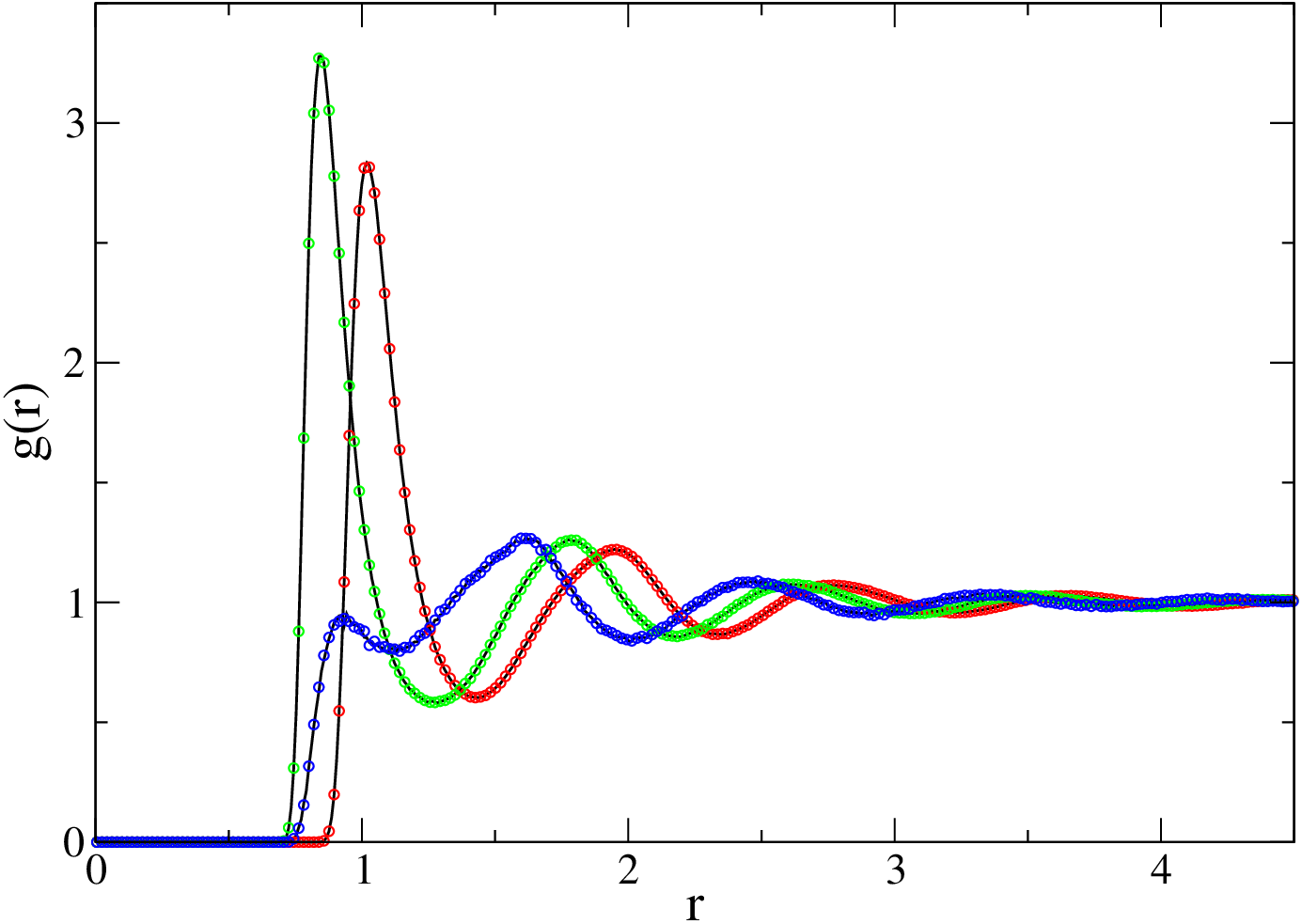}
    \includegraphics[width=70mm]{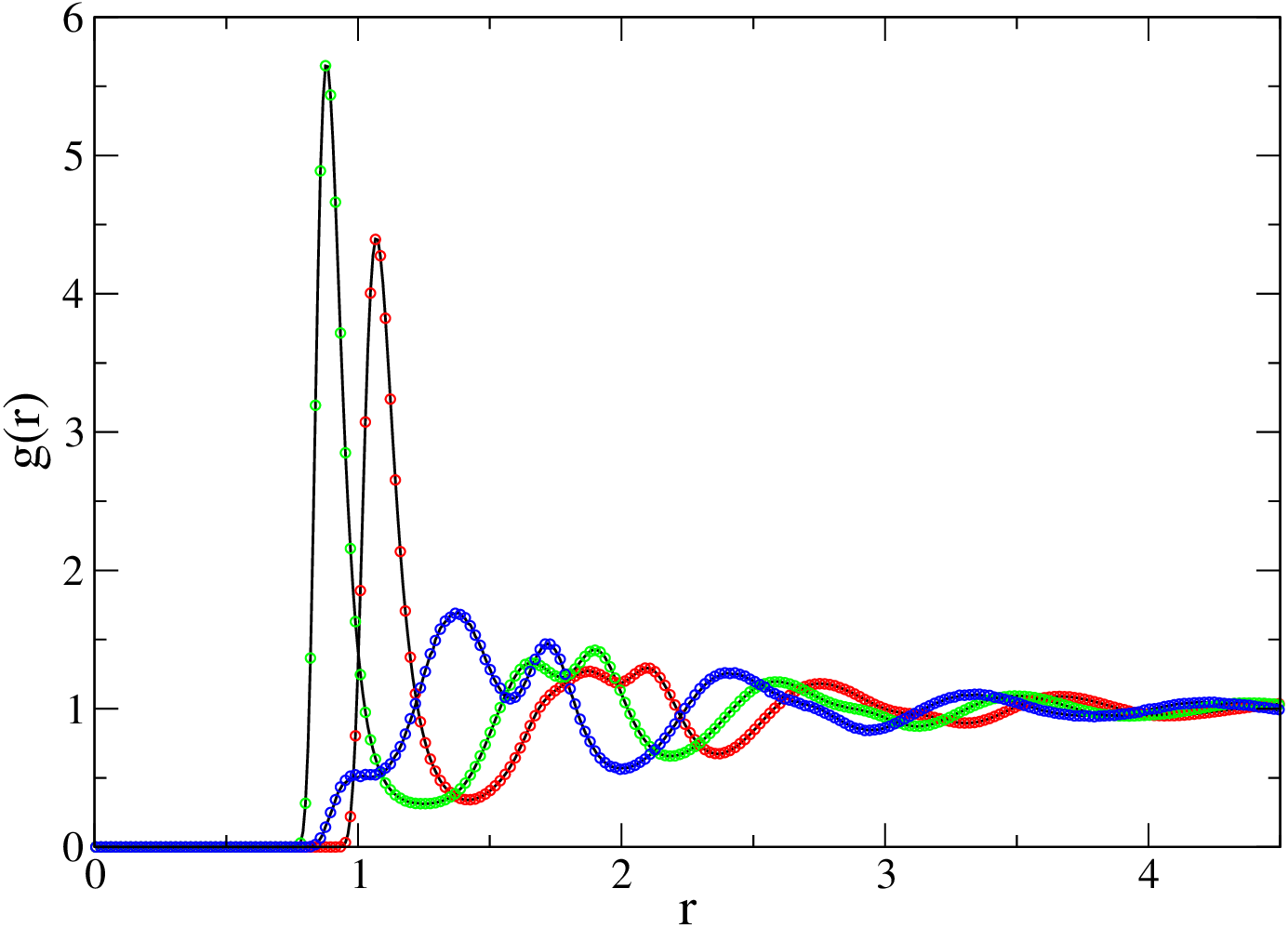}
  \caption{\label{rdfka}The radial distribution functions for the KABLJ system at $\rho = 1.2$. The black lines give results from $NVE$ simulations, colored circles from $NVU$ simulation where green, red, and blue denote, respectively, AB, AA, and BB pairs.
(a) $T=2.0$; 
(b) $T=0.405$.
}
\end{figure}

In view of the above, for comparing $NVU$ and $NVE$ generated sequences we define the $NVU$ time step length $\dt_{NVU}$ as the average of Eq. (\ref{time_step}), i.e.,

\begin{equation}\label{time}
\frac{(\dt_{NVU})^2}{m}\,\equiv\,  
-2\,\left\langle{\frac{\bF_{i} \cdot (\bR_{i} - \bR_{i-1})}{\bF_{i}^{2}}}\right\rangle\,.
\end{equation}

\begin{figure}[H]
\includegraphics[width=70mm]{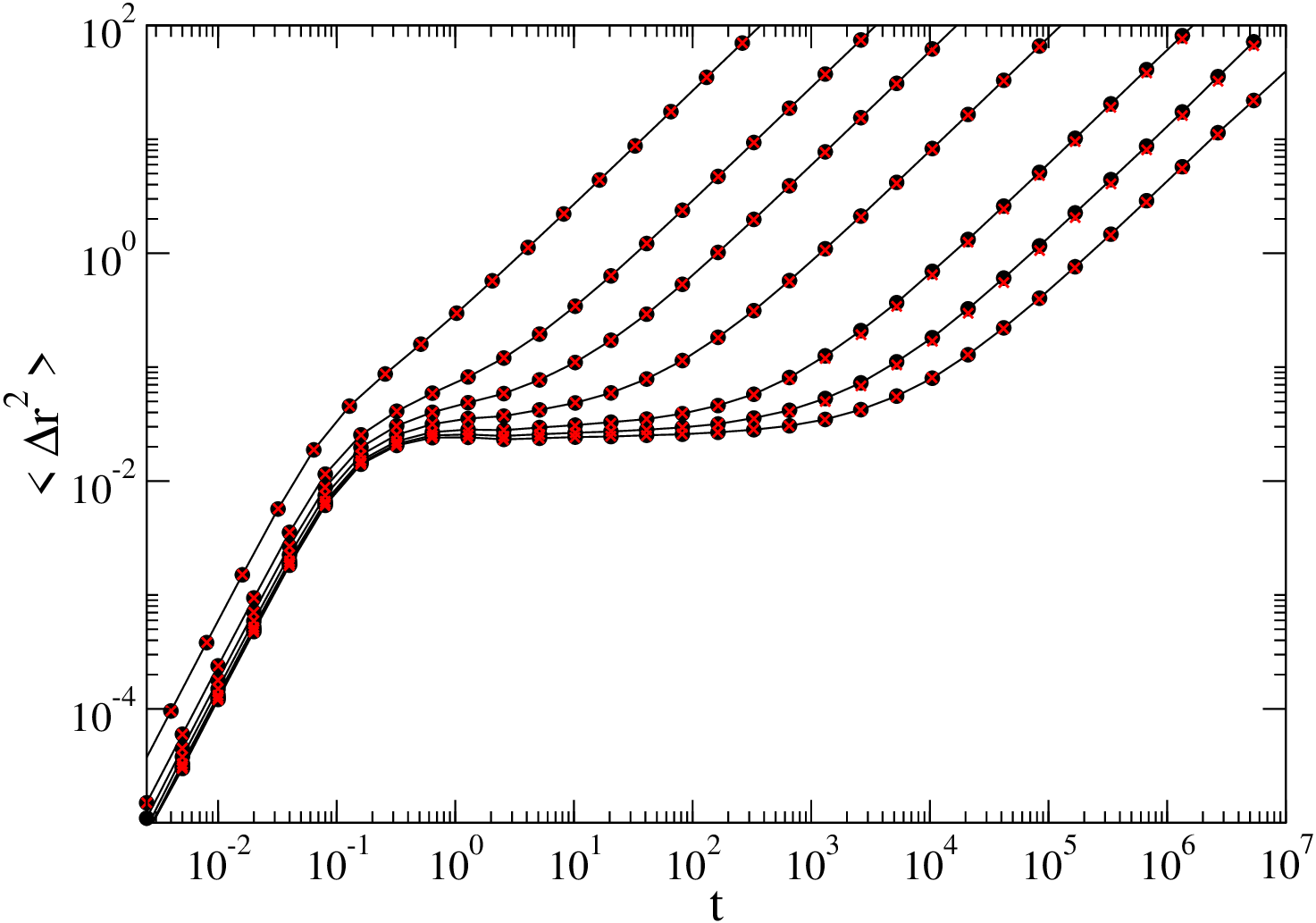}
 \includegraphics[width=70mm]{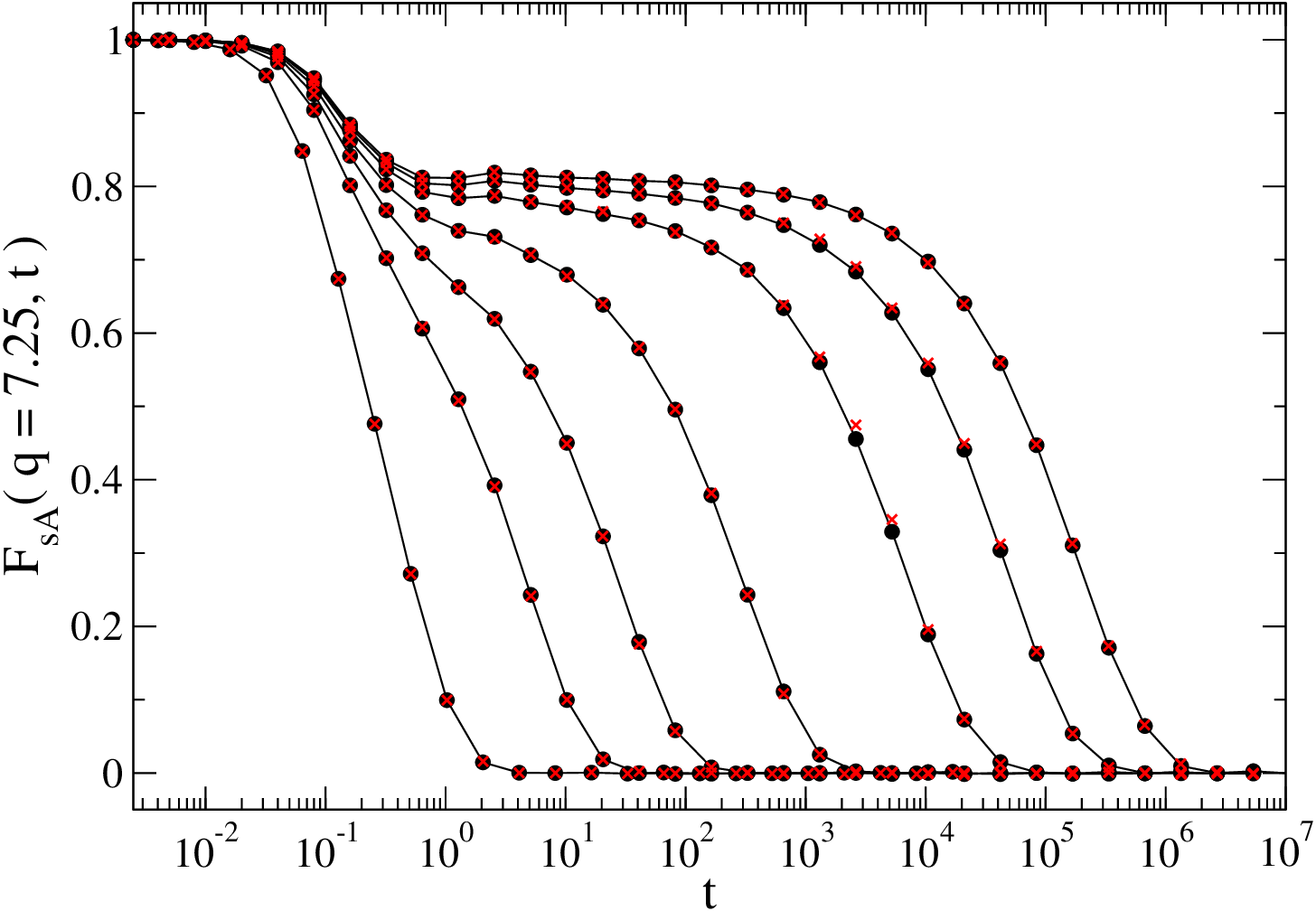}
    \caption{\label{msdfska}(a) Mean-square displacement and (b) incoherent intermediate scattering function at the wave vector of the first peak of the AA structure factor. Both simulations were performed at $\rho = 1.2$ and  $T = 2.0,\, 0.80,\, 0.60,\, 0.50,\, 0.44,\,0.42,\,0.405$ (left to right) for the KABLJ mixture (1024 particles). $NVE$ dynamics is given by the filled black circles connected by straight lines, $NVU$ dynamics by the red crosses.}
\end{figure}

First, we present results that compare static averages of $NVU$ and $NVE$ simulations. Figure \ref{rdfka} shows the three radial distribution functions for the KABLJ liquid at two different state points. Clearly the two algorithms give identical results. Figure \ref{msdfska} shows $NVU$ and $NVE$ results for the mean-square displacement and the incoherent intermediate scattering function of the KABLJ liquid at density $\rho = 1.2$ over a range of temperatures. The mean-square displacement and the incoherent scattering function are both identical for $NVU$ and $NVE$ dynamics.

\begin{figure}[H]
 \includegraphics[width=70mm]{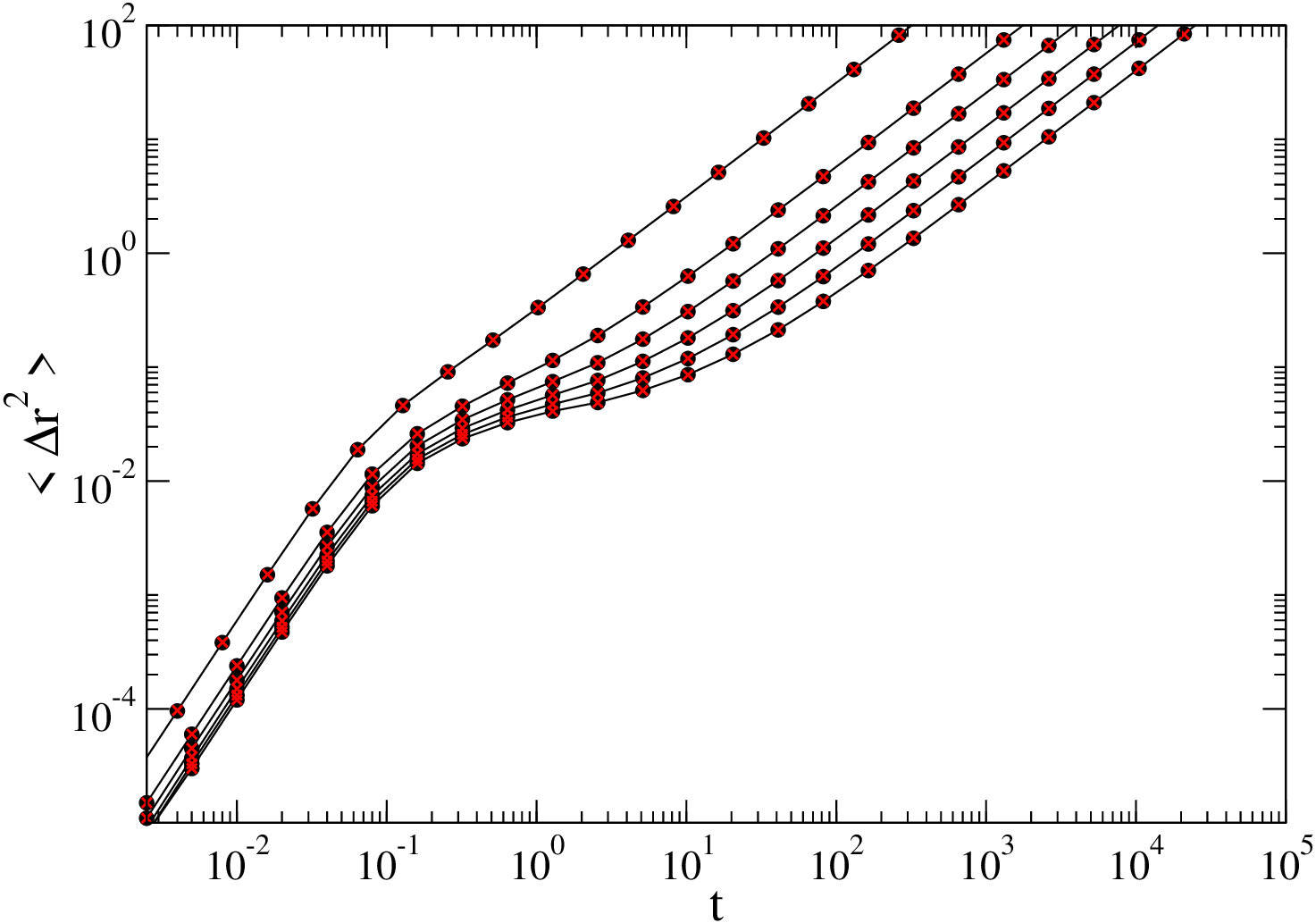}
    \includegraphics[width=70mm]{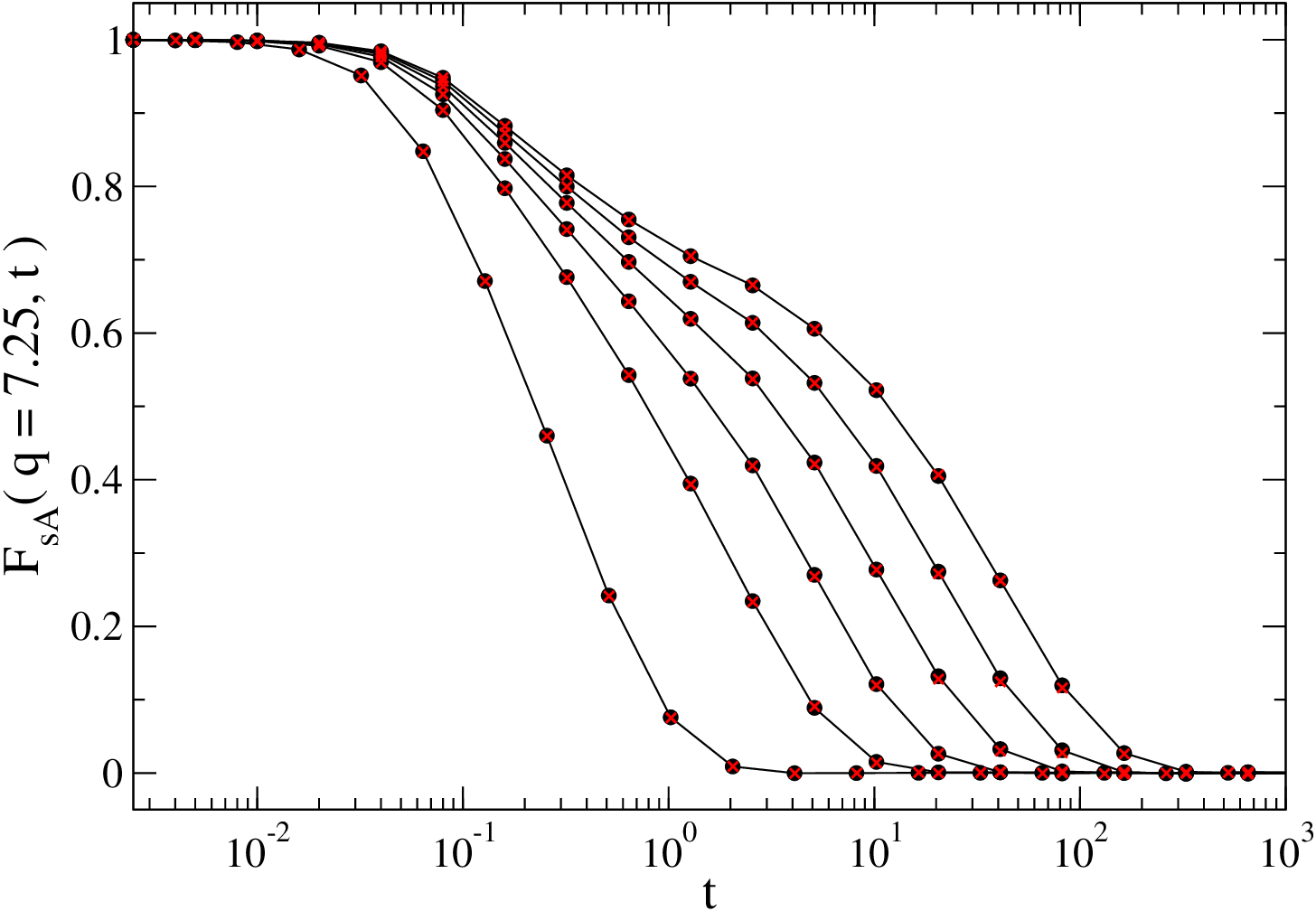}
  \caption{\label{msdfswca}(a) Mean-square displacement and (b) incoherent intermediate scattering function at the same wave vector as in Fig. \ref{msdfska}. Both simulations were performed at $\rho = 1.2$ and $T = 2.0,\, 0.80,\, 0.60,\, 0.50,\, 0.44$ and $0.40$ (left to right) for the WCA approximation to the KABLJ mixture. $NVE$ dynamics is given by the filled black circles connected by straight lines, $NVU$ dynamics by the red crosses.}
\end{figure}

Corresponding figures are shown in Fig. \ref{msdfswca} for the Weeks-Chandler-Andersen (WCA) approximation, which cuts off interactions beyond the energy minima, i.e., keep only the repulsive part of the potential. The WCA version of the system has a similar structure, but a much faster dynamics in the supercooled regime \cite{scldynamics,scl}. Again, $NVU$ and $NVE$ dynamics give identical results.

\begin{figure}[H]
  \centering
  \includegraphics[width=70mm]{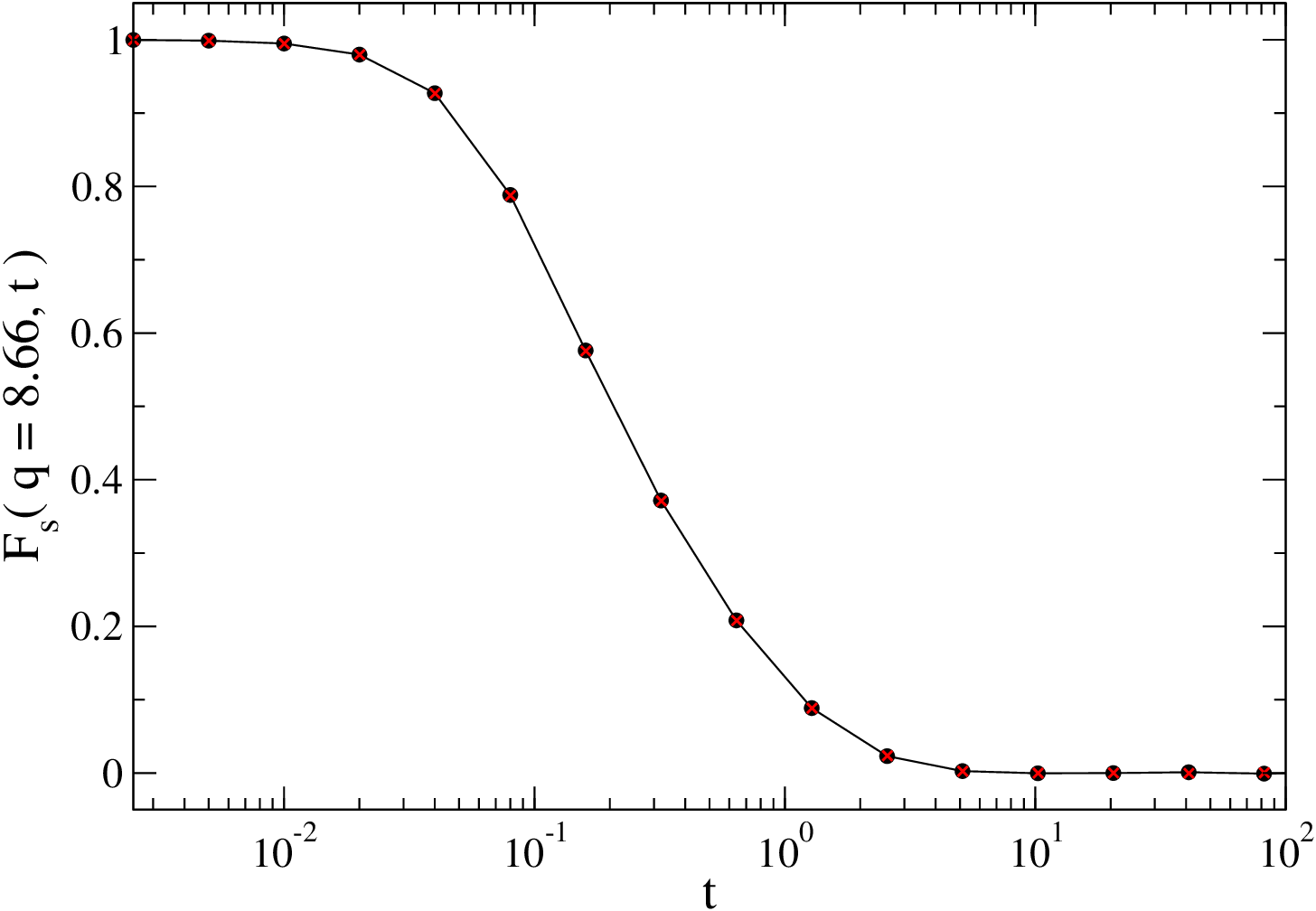}
  \caption{The incoherent intermediate scattering function at $\rho = 0.8$ and $T = 1.4$ for a Lennard-Jones Gaussian system \cite{ljg}. The black circles represent an $NVE$ simulation, the red symbols an $NVU$ simulation. It should be noted that the system phase separated during the simulation.}
  \label{gauss}
\end{figure}

We studied also the so-called Lennard-Jones Gaussian system defined by a pair potential that adds a Gaussian to a LJ potential \cite{ljg}, a liquid that is not strongly correlating \cite{scl}. Figure \ref{gauss} shows that also for this model the incoherent intermediate scattering function is the same for $NVU$ and $NVE$ dynamics. In conclusion, for all systems simulated we found $NVU=NVE$. This applies even for $N=65$ particles of the KABLJ liquid ($T=0.8$, $\rho=1.2$).

\section{Intuitive arguments for the equivalence of $NVU$ and $NVE$ dynamics as $N\rightarrow\infty$}\label{proof}

The above results raise the question: Are $NVU$ and $NVE$ dynamics mathematically equivalent in some well-defined sense? The two algorithms are not identical, of course; that would require no variation in the quantity $\dt_{i,NVU}$ (Fig. \ref{ldist}). On the other hand, the $\dt_{i,NVU}$ distribution narrows as the particle number increases (Fig. \ref{ldist} (b)). From this $NVU$ and $NVE$ dynamics are expected to become equivalent for $N\rightarrow\infty$ in the following sense: For any configurational quantity $A$, to leading order in $1/N$ there is identity of dynamic quantities like the time-autocorrelation function $\langle A(0)A(t)\rangle$ or the mean-square change $\langle \Delta^2 A(t)\rangle$ (i.e., the relative deviations go to zero as $N\rightarrow\infty$). Consider the time-autocorrelation function of an extensive quantity $A$ with zero average. In this case the time-autocorrelation function scales in both ensembles as $N$, and the proposed equivalence of the dynamics means that $|\langle A(0)A(t)\rangle_{NVU}-\langle A(0)A(t)\rangle_{NVE}|\propto N^0$ as $N\rightarrow\infty$. Intuitively, what happens is that, since in $NVE$ dynamics the relative potential-energy fluctuations go to zero $N\rightarrow\infty$, it becomes a better and better approximation to regard the potential energy as conserved \cite{stratt}.

There exists in analytical mechanics a variational principle that does not involve time at all. This is the Maupertuis principle from 1746  \cite{lan,wiki}, a variational principle that is originally due to Jacobi and for this reason is sometimes referred to as ``Jacobi's form of the least action principle'' \cite{wiki,anal_mech}. This states that a classical-mechanical system of fixed energy $E$ follows a curve in configuration space obeying (with fixed end points)

\begin{equation}\label{maup}
  \delta \int_{\bR_A}^{\bR_B} \sqrt{2m(E-U)}\,dl\,\,=\, 0\,.
\end{equation}
One may argue that the relative variations of the integrand go to zero as $N\rightarrow\infty$. Thus the integrand in this limit becomes effectively constant and can be taken outside the variation, implying Eq. (\ref{cond}) for motion that effectively takes place on the constant potential energy surface \cite{stratt}.

If $l$ is the path length parametrizing the path, Eq. (\ref{maup}) implies \cite{anal_mech,lan} $d^2\bR/dl^2=[\bF - (\bF\cdot\bt)\bt]/2(E-U(\bR))$ where $\bt=d\bR/dl$ is the unit vector tangential to the path. The term $\bF - (\bF\cdot\bt)\bt$ is the (vector) component of the force normal to the path. In the thermodynamic limit the path approaches the constant-potential-energy hypersurface $\Omega$, i.e., $\bF\cdot\bt=0$. In this limit one has $dl\propto dt$ because the relative kinetic energy fluctuations go to zero. In this way the Maupertuis principle is equivalent to both the geodesic equation Eq. (\ref{cond}) and to Newton's second law $\ddot{\bR} = \bF/m$.

\begin{figure}[H]
  \centering
  \includegraphics[width=70mm]{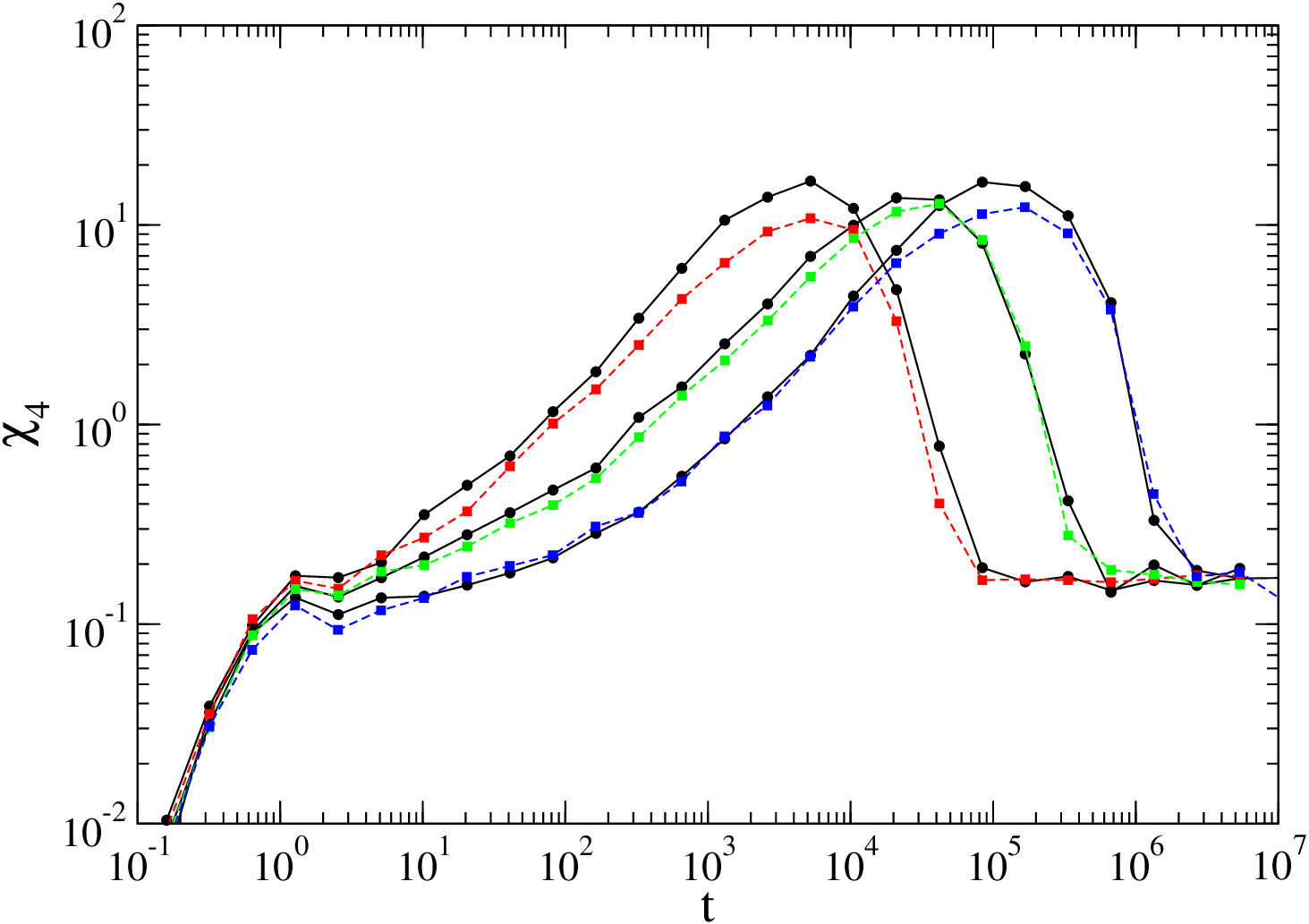}
  \includegraphics[width=70mm]{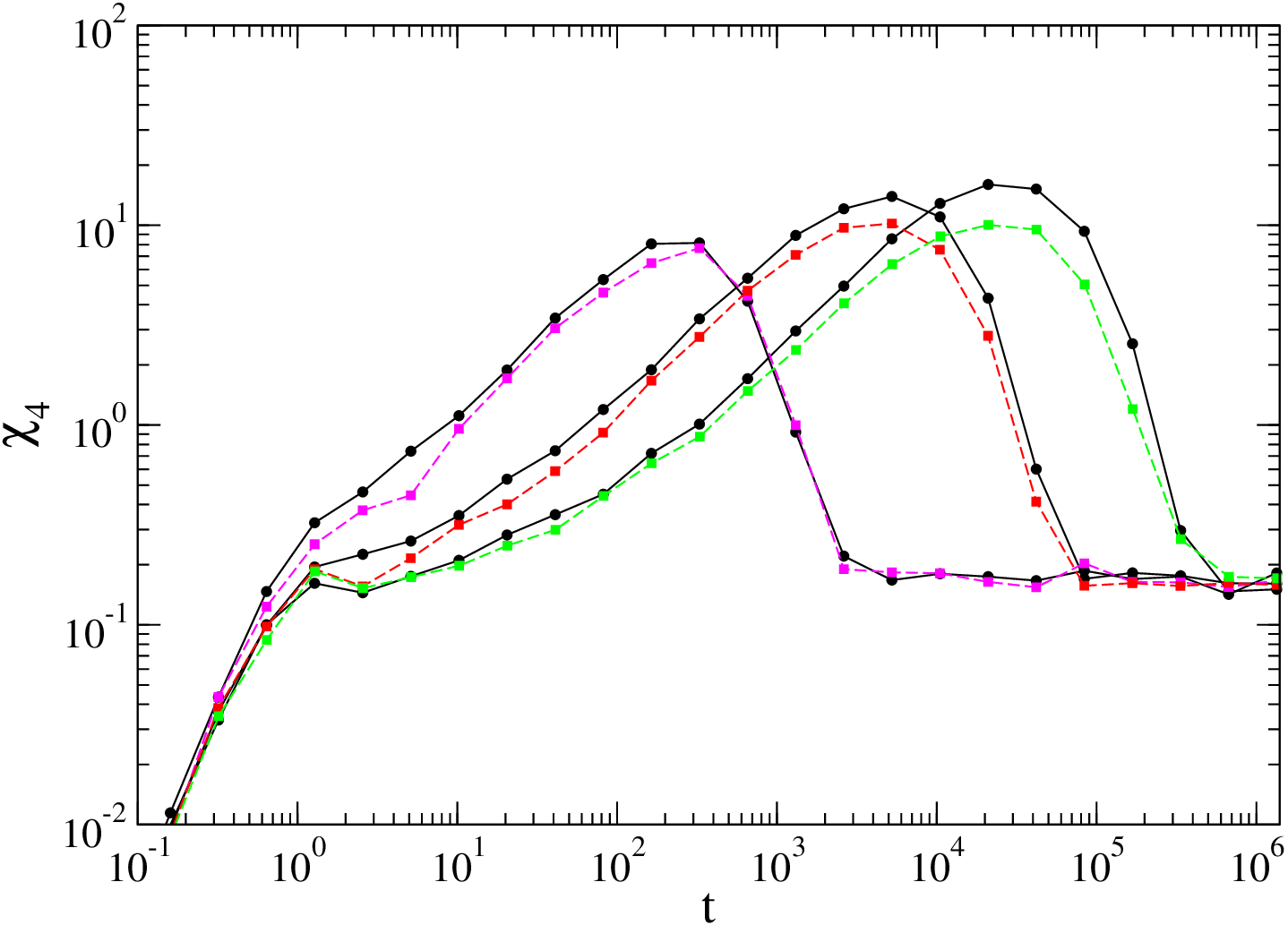}
  \caption{\label{chi4}
(a) The dynamical fluctuations quantified by $\chi_{4}(t)$ for the $A$ particles at $\rho = 1.2$ for a system with 1024 particles. The black circles give results for an $NVE$ simulation, the red, green, and blue symbols represent $NVU$ simulations at, respectively, $T=0.44,\,0.42,\,0.405$.
(b) The dynamical fluctuations quantified by $\chi_{4}(t)$ for the $A$ particles at $\rho = 1.2$ for a system with 2048 particles. The black circles give results for an $NVE$ simulation, the violet and red symbols represent $NVU$ simulations of, respectively, $T=0.50,\,0.44,\,0.42$. Increasing the number of particles does not appear to decrease the deviation between the two dynamics.
}
  \end{figure}

The equivalence of $NVU$ and $NVE$ dynamics in the thermodynamic limit relates to static averages as well as to time-autocorrelation functions of extensive quantities. Just as one must be careful when comparing fluctuations between different ensembles, fluctuations relating to the dynamics need not be the same for $NVU$ and $NVE$ dynamics. As an example, Fig. \ref{chi4} shows the quantity $\chi_{4}(t)$ defined by $\chi_{4}(t) = N_{A}\,[\,\langle F_{sA}^{2}(\textbf{k}, t)\rangle - \langle F_{sA}(\textbf{k}, t)\rangle^{2}\,]\,$ for the KABLJ system at three temperatures and two values of $N$. $\chi_{4}$ quantifies the incoherent intermediate scattering function fluctuations \cite{chi4}. For $\chi_{4}(t)$ $NVU$ and $NVE$ dynamics do not appear to give identical results. A related observation was made by Berthier \textit{et al.}, who showed that $\chi_{4}(t)$ is not the same in $NVE$ and in $NVT$ dynamics \cite{berthierMC}.

\section{Comparing $NVU$ dynamics to $NVT$, Monte Carlo, and diffusion on $\Omega$}\label{nvuother}

This section compares simulations using $NVU$ dynamics to results for three further dynamics, two of which are standard. We focus on the viscous regime. One dynamics is the Nos$\acute{e}$-Hoover \textit{NVT} dynamics, a deterministic sampling of the $NVT$ canonical ensemble that may be derived from a ``virtual'' Hamiltonian \cite{nosehoovernvt, algonvt}. The second standard dynamics considered is the Metropolis Monte Carlo (MC) algorithm, which generates a stochastic sequence of states giving the correct $NVT$ canonical ensemble distribution. The third dynamics employed below is also stochastic, it simulates diffusion on the constant-potential-energy hypersurface $\Omega$ by a small step-length random walk (RW) on $\Omega$. This was discussed by Scala {\it et al.} \cite{scala}, who proposed the following equation of motion

\begin{equation}\label{rw}
  \frac{d\bR_{i}}{dt} = \Delta {\bm\eta}_{i} - \frac{\Delta {\bm\eta}_{i} \cdot \textbf{F}_{i}}{\textbf{F}_{i}^{2}} \textbf{F}_{i}\,,
\end{equation}
where $\Delta {\bm\eta}_{i}$ is a $3N$-dimensional random vector (see below). Equation (\ref{rw}) implies $\bF_{i} \cdot \dot{\bR}_{i} = 0$, which ensures the potential-energy conservation required for staying on $\Omega$. 

The RW algorithm was discretized and implemented as a ''predictor-corrector'' algorithm in the following way. A vector $\Delta {\bm\eta}_{i}$ was chosen from a  cube with length $L = 0.01 \sigma$. This is small enough to ensure that the dynamics generates the correct $NVE$ radial distribution function and at the same time has no effect on the average dynamical quantities. Positions were updated via

\begin{equation}
  \textbf{R}_{i+1} = \bR_{i} + \Delta t\Delta {\bm\eta}_{i} - \frac{\Delta t\Delta {\bm\eta}_{i} \cdot \textbf{F}_{i}}{\textbf{F}_{i}^{2}} \textbf{F}_{i}\,.
\end{equation}
Finally, $\textbf{R}_{i+1}$ was corrected by applying two iterations of $\bR_{i+1} \equiv \bR_{i+1} - \frac{U_{i+1} - U_{0}}{\textbf{F}_{i+1}^{2}} \textbf{F}_{i+1}$ in order to eliminate long-time entropic drift of the potential energy.

$MC$ and $RW$ dynamics involve no generic measures of time. We compared their results to $NVU$ dynamics by proceeding as follows. At any given state point the time-scaling factor was determined from the long-time behavior of the mean-square displacement by requiring that the single-particle displacement obeys $\langle\Delta x^2(t)\rangle=2Dt$ for $t\rightarrow\infty$ with the $NVE$ diffusion constant $D$. By construction this ensures agreement with the long-time mean-square displacement of $NVE$ dynamics.

\begin{figure}[H]
  \includegraphics[width=70mm]{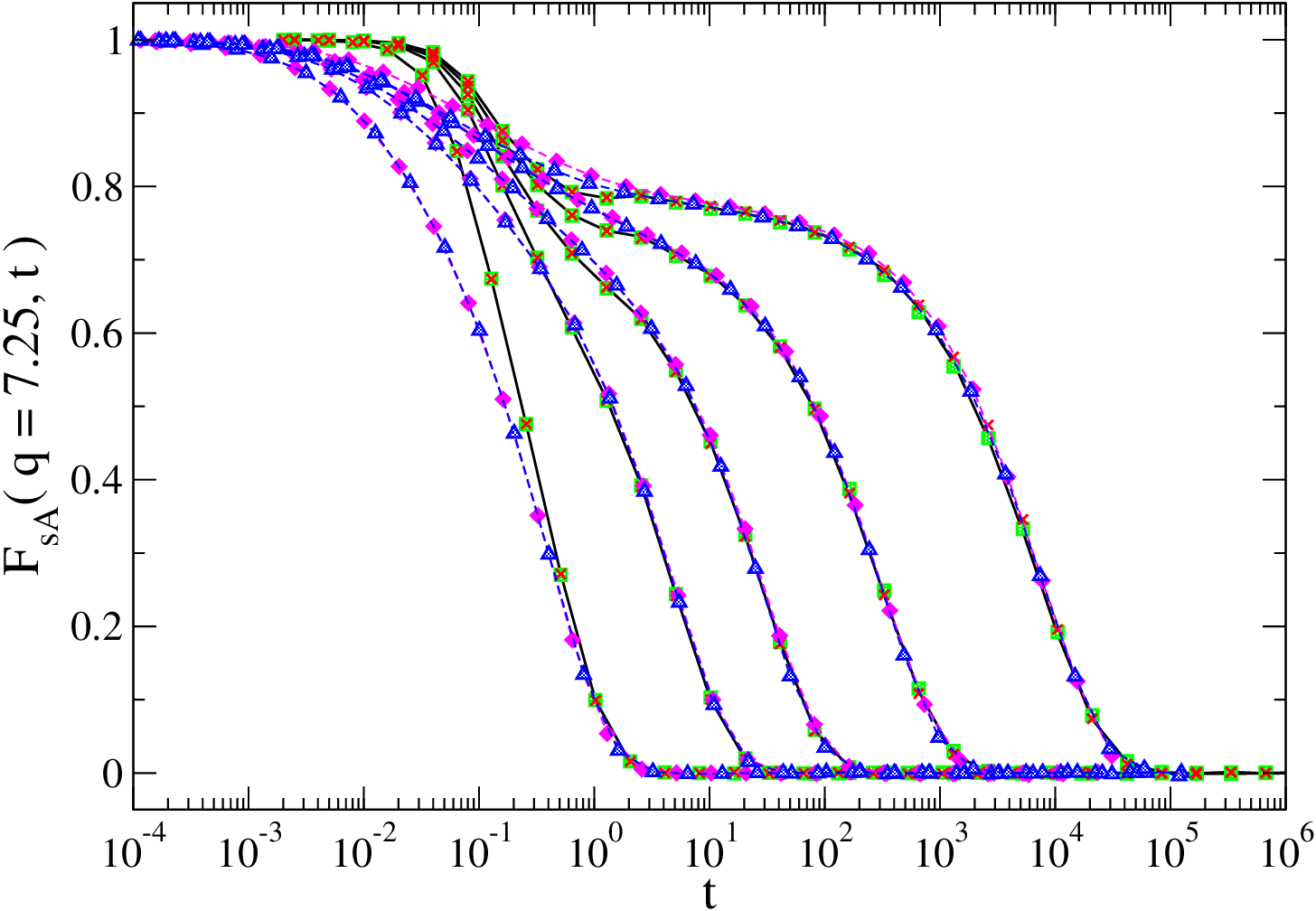}
  \centering
  \caption{\label{kafsall}The incoherent intermediate scattering function for all five investigated dynamics for the KABLJ mixture at $\rho = 1.2$ and $T=2.0,\, 0.80,\, 0.60,\, 0.50$ and $0.44$. The black curve (with filled circles) is the $NVE$ simulation, red crosses: $NVU$, green squares: $NVT$, magenta diamonds: $MC$, blue triangles: $RW$.}
\end{figure}

In Fig. \ref{kafsall} we show the incoherent intermediate scattering function of the KABLJ liquid for all investigated dynamics at several state points.

$NVU$ and $NVT$ dynamics agree quantitatively for all investigated state points. This is not surprising given the results of Secs. \ref{nvunve} and \ref{proof} and the well-known fact that $NVE$ and $NVT$ dynamics give same time-autocorrelation functions to leading order in $1/N$ \cite{evans}. The incoherent intermediate scattering functions of $MC$ and $RW$ agree at all investigated temperatures. This is consistent with the recent results of Berthier \textit{et al.} \cite{berthierMC}, who compared Langevin to \textit{MC} dynamics. For lower temperatures ($T < 0.80$) quantitative agreement is found among all five dynamics investigated in the $\alpha$-relaxation regime.

\begin{figure}[H]
    \includegraphics[width=70mm]{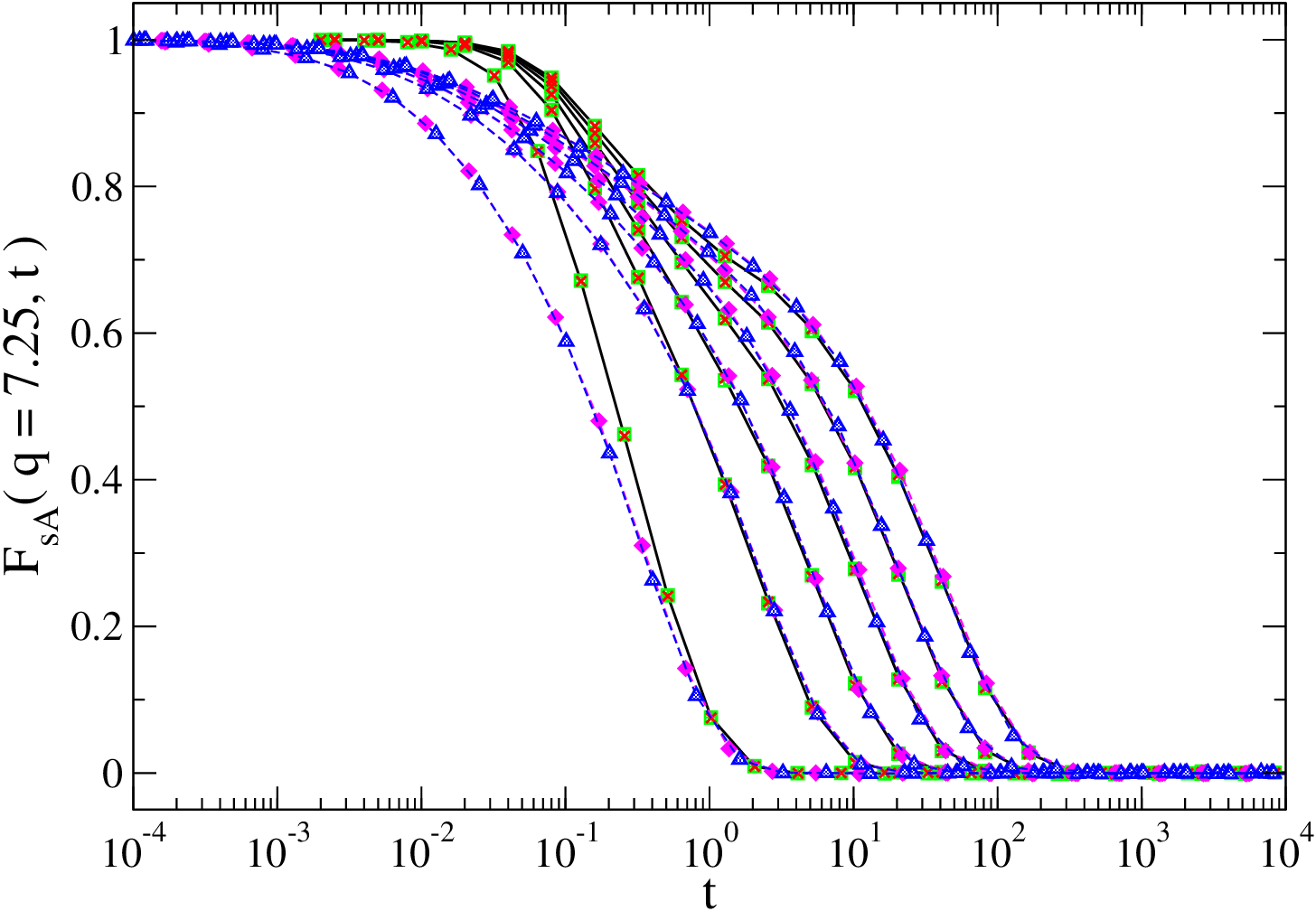}
    \centering
    \caption{The incoherent intermediate scattering function for all five investigated dynamics for the KABWCA system at $\rho = 1.2$ and $T=2.0,\, 0.80,\, 0.60,\, 0.50,\, 0.44$ and $0.40$. The black curve (with filled circles) is the $NVE$ simulation, red crosses: $NVU$, green squares: $NVT$, magenta diamonds: $MC$, blue triangles: $RW$.}
  \label{wcafsall}
\end{figure}

The corresponding figure for the KABWCA system is shown in Fig. \ref{wcafsall}. The same conclusion is reached as for the KABLJ mixture.

\section{Summary and outlook}\label{sum}

$NVU$ dynamics traces out geodesic curves on the $3N-1$ dimensional potential-energy hypersurface $\Omega$. We have compared $NVU$ dynamics with other dynamics. Simulations supplemented by non-rigorous analytical arguments showed that $NVU$ and $NVE$ dynamics are equivalent in the thermodynamic limit, i.e., typical autocorrelation functions become identical to leading order in $1/N$. Furthermore, $NVU$ dynamics was compared to two stochastic dynamics, standard Monte Carlo dynamics and a small-step random walk on the constant-potential-energy hypersurface $\Omega$ representing diffusion on $\Omega$. Agreement was established for all dynamics, including also $NVT$ dynamics, in the $\alpha$-relaxation regime where inertial effects become unimportant. We conclude that $NVU$ dynamics is a fully valid molecular dynamics. 

It is interesting to note that $NVU$ dynamics, like any geodesic motion on a Riemannian manifold, can be formulated as a Hamiltonian dynamics based on the curved-space purely kinetic energy Hamiltonian $H=1/2\sum_{a.b} g_{ab}(x)p^a p^b$ where $x$ is the manifold coordinate, $g_{ab}$ the corresponding metric tensor, and $p_a$ the generalized momenta \cite{wikip}. Indeed, long ago Hertz argued that one should focus exclusively on the kinetic energy and describe classical mechanics as a geodesic motion on a high-dimensional Riemannian manifold (along the ``geradeste Bahn'' of this manifold, the straightest curve) \cite{Hertz}. Hertz' idea was to eliminate the force and potential energy concepts entirely from mechanics and replace particle interactions by constraints among the coordinates; the relevant manifold is defined by these constraints. This is not what we have done here. There is, however, the fundamental similarity between the Hertz and the $NVU$ approaches that both build on the conceptual simplification of  ``replacing Newton's second law by Newton's first law''. Moreover, as shown in Appendix A the effect of masses enters into the metric of the Riemannian manifold in precisely the same way as we need for $NVU$ dynamics when it is generalized to deal with systems of varying masses. Thus $NVU$ dynamics realizes Hertz's ideas to a large extent.

From a technical point of view $NVU$ dynamics provides few advantages because it is not faster than $NVE$ or $NVT$ dynamics. However, by referring directly to the mathematical properties of a Riemannian differentiable manifold $NVU$ dynamics leads to a new way of thinking about the classical mechanics of many-particle systems. Future work should focus on relating the mathematical properties of $\Omega$ to the physical properties of the system in question. It is our hope that in this way new insights into liquid dynamics may be arrived at by adopting the $NVU$ viewpoint.

\acknowledgments 
Useful inputs from Ole J. Heilmann and Nick Bailey are gratefully acknowledged. The centre for viscous liquid dynamics ``Glass and Time'' is sponsored by the Danish National Research Foundation (DNRF).

\appendix

\section{Generalization of the $NVU$ algorithm to deal with systems of different particles masses}

Papers I and II deal with systems of particles with identical mass $m$. The basic $NVU$ algorithm Eq. (\ref{numgeo}), however, is well defined and works perfectly well for any general classical mechanical system. The algorithm traces out a geodesic on $\Omega$ that is independent of the particles' masses, a geometrical path entirely determined from the function $U(\br_1,...,\br_N)$. Equation (\ref{time_step}), which ensures $NVU=NVE$ in the thermodynamic limit, only works if all particles have mass $m$. The question arises if a generalization of Eq. (\ref{numgeo}) is possible ensuring that $NVU=NVE$ as $N\rightarrow \infty$ also for systems of particles with different masses.

If the $k$'th particle mass is $m_k$, we seek to modify the basic $NVU$ algorithm such that it for the $k$'th particle as $N\rightarrow \infty$ converges to (where $\br^{(k)}$ is the coordinate of the $k$'th particle, $\bF^{(k)}$ the force on it, and subscript $j$ is the time step index)

\begin{equation}\label{verlet_k}
\br_{j+1}^{(k)} \,=\, 
2\,\br_j^{(k)} - \br_{j-1}^{(k)} +\frac{(\dt)^2}{m_k} \bF_j^{(k)}\,.
\end{equation}
If the average mass is $\am$, we define reduced masses by

\begin{equation}\label{red_mass}
\tilde m _k\,\equiv\,
\frac{m_k}{\am}\,.
\end{equation}
A geodesic is defined by giving the shortest distance between any two of its close by points. In Paper I and in Eq. (\ref{cond}) of the present paper the distance measure is given by the standard Euclidian distance $dl^2=\sum_k d\br^{(k)}\cdot d\br^{(k)}$. A change of the metric leads to different geodesics. Consider the following metric:

\begin{equation}\label{metric}
dl^2\,=\,
\sum_k \tilde m _k \,d\br^{(k)}\cdot d\br^{(k)}
\end{equation}
This is precisely the metric discussed by Hertz in his mechanics long ago \cite{Hertz}. In the ``Hertzian'' metric the discritezed path length used in deriving the $NVU$ algorithm is (Paper I) $\sum_j\sqrt{\sum_k \tilde m _k \left(\br_j^{(k)}-\br_{j-1}^{(k)}\right)^2}$ (where $j$ is the time step index). Thus the variational condition becomes

\begin{equation}\label{nvu_cond}
\delta \left(\sum_j \sqrt{\sum_k \tilde m _k \left(\br_j^{(k)}-\br_{j-1}^{(k)}\right)^2} -\sum_j\lambda_j U( \bR_j)\right)\,=\,0\,.
\end{equation}
From this it follows via the ansatz of constant step length that

\begin{equation}\label{nvunew1}
\br_{j+1}^{(k)}\,=\,
2\br_j^{(k)}-\br_{j-1}^{(k)}-2 [\bF_j\cdot\left( \bR_j-\bR_{j-1}\right)]\bF_j^{(k)}/(\tilde m _k\bF_j^2)\,.
\end{equation}
This translates into Eq. (\ref{verlet_k}) for a suitably chosen $\Delta t$; likewise the relative fluctuations of the term $2 [\bF_j\cdot\left( \bR_j-\bR_{j-1}\right)]/\bF_j^2$ go to zero in the thermodynamic limit ($N\rightarrow\infty$), such that $NVU=NVE$ in this limit. 

It is important to note that when systems of varying masses are considered, there is also the option of ignoring the masses as done by Stratt and coworkers \cite{stratt}. This approach, which is consistent with, e.g., Brownian mechanics, leads to a perfectly admissible $NVU$ dynamics. However, it does not correspond to $NVE$ dynamics in the same way as the above proposed ``Hertzian'' version of $NVU$ dynamics with varying masses.

\end{document}